\documentclass[aps,prb,twocolumn]{revtex4}
\usepackage{amsmath}
\usepackage{graphicx}
\usepackage{color}

\newcommand{\be}{\begin{equation}}
\newcommand{\ee}{\end{equation}}
\newcommand{\bea}{\begin{eqnarray}}
\newcommand{\eea}{\end{eqnarray}}

\newcommand{\lb}{\left[}
\newcommand{\rb}{\right]}
\newcommand{\lp}{\left(}
\newcommand{\rp}{\right)}

\renewcommand{\epsilon}{\varepsilon}
\renewcommand{\Im}{{\rm Im}\,}

\renewcommand{\vec}[1]{{\bf #1}}

\begin{document}
\title{Coulomb Drag Mechanisms in Graphene}
\author{J. C. W. Song${}^{1,2}$}
\author{D. A. Abanin${}^{3}$}
\author{L. S. Levitov${}^{1}$}
\affiliation{$^1$ Department of Physics, Massachusetts Institute of Technology, Cambridge, MA 02139}
\affiliation{$^2$ School of Engineering and Applied Sciences, Harvard University, Cambridge, Massachusetts 02138, USA}
\affiliation{$^3$ Perimeter Institute for Theoretical Physics, Waterloo, Ontario N2L 6B9, Canada}
\date{\today}
\begin{abstract}
Recent measurements revealed an anomalous Coulomb drag in graphene, hinting at new physics at charge neutrality. The anomalous drag is explained by a new mechanism based on energy transport, which involves interlayer energy transfer, coupled to charge flow via lateral heat currents and thermopower. The old and new drag mechanisms are governed by distinct physical effects, resulting in starkly different behavior, in particular for drag magnitude and sign near charge neutrality. The new mechanism explains the giant enhancement of drag near charge neutrality, as well as its sign and anomalous sensitivity to magnetic field. Under realistic conditions,  energy transport dominates in a wide temperature range, giving rise to a universal value of drag which is essentially independent of the electron-electron interaction strength.
\end{abstract}

\maketitle

\section{Introduction}

Long-range Coulomb interactions have long been known to result in a kind of spooky action between adjacent electrically isolated electron systems arising when current applied in  one (active) layer induces voltage in the second (passive) layer (Fig. 1a). Such phenomena, known as Coulomb drag, occur despite the lack of particle exchange between layers and provides one of the most sensitive probes of interactions in low-dimensional systems.
Coulomb drag was extensively studied in GaAs quantum wells,\cite{gramila,sivan} where the observations were successfully interpreted in terms of the momentum drag mechanism\cite{jauho,zheng,kamenev,flensberg} (hereafter referred to as ``P-mechanism''), in which interlayer electron-electron scattering  mediated by long-range Coulomb interaction transfers momentum from the active layer to the passive layer.

Recent measurements of Coulomb drag in double layer graphene heterostructures\cite{tutuc,geim} revealed  strong drag with unusual density dependence near charge neutrality. This behavior was attributed to
the close proximity of graphene layers, with 
 typical layer separation reaching values as small as 1-2 nm.\cite{geim} Importantly, this is much smaller than characteristic lengthscales such as the electron Fermi wavelength and the screening length, and also much smaller than layer separations in 
 GaAs-based structures. This defines a new strong-coupling regime wherein the interlayer and intralayer interactions are nearly equally strong. Fast momentum transfer between electron subsystems in the two layers
and strong Coulomb drag have been predicted in this regime.\cite{tse2007,sensarma2010,narozhny2007,peres2011,katsnelson2011,narozhny2}

Beyond sheer enhancement of drag, measurements in G/hBN/G heterostructures\cite{geim} unveiled puzzling new features close to the double neutrality point (DNP). Ref. \onlinecite{geim} found that, in contrast to predictions from P-mechanism,
drag did not vanish at DNP. Instead, drag resistivity peaked at DNP where it exhibited anomalous sensitivity to magnetic field, becoming {\it colossal} (increasing by more than a factor of $10$) and reversing sign from positive to {\it negative} when a classically weak field as low as $B=0.2 \, {\rm T}$ was applied\cite{geim}.

\begin{widetext}

\begin{figure}
\includegraphics[scale=0.25]{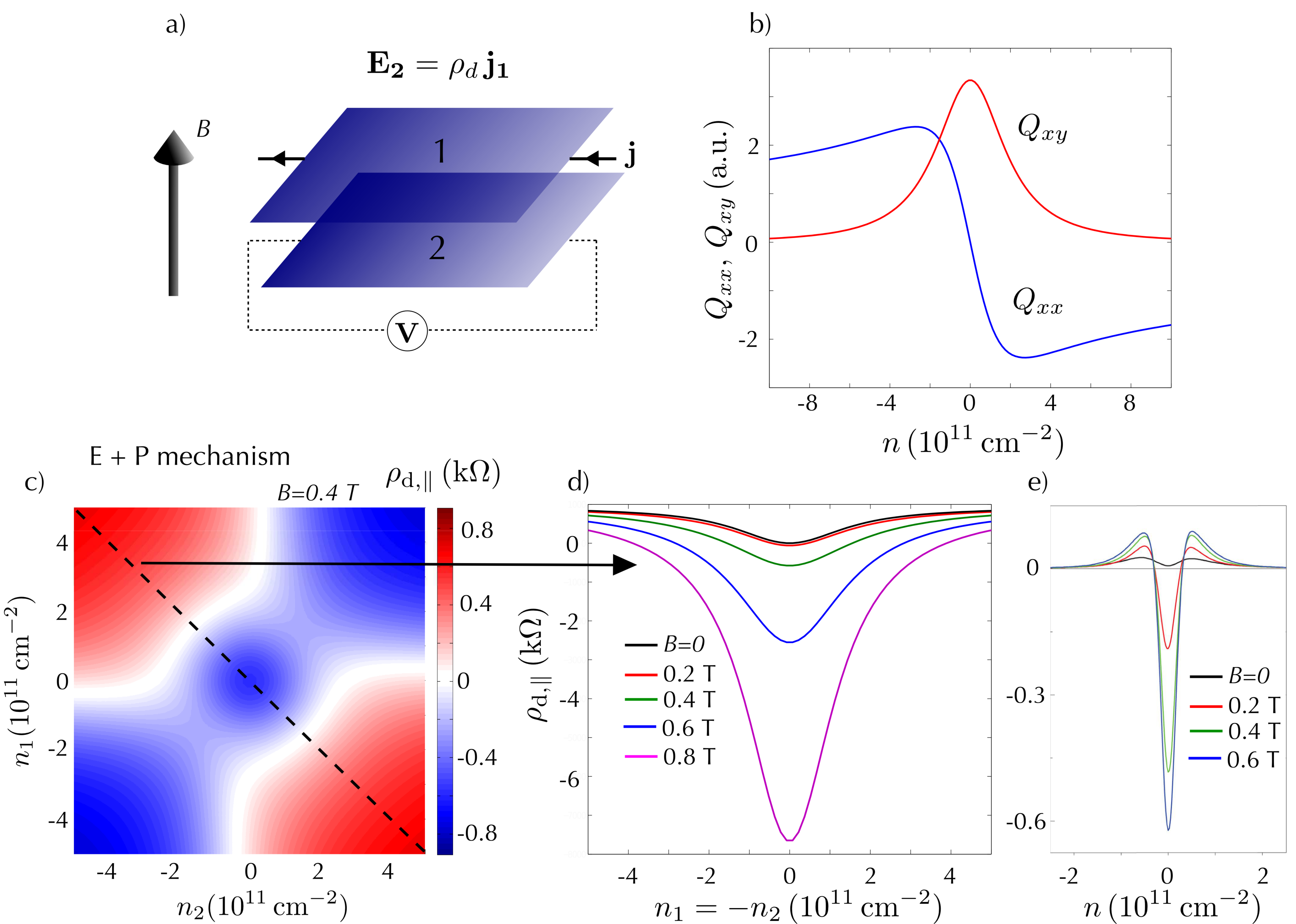}
\caption{Two adjacent layers of graphene can exhibit a drag resistivity, $\rho_{\rm d}$, that features giant enhancement near the double neutrality point (DNP). (a) Schematic of double-layer graphene heterostructure and drag measurement. (b) Thermoelectric coupling $Q_{xx} = \rm{Re}(Q)$, $Q_{xy} = - {\rm Im}(Q)$, Eq.(\ref{eq:Q}),  peaks close to the Dirac point.
(b) Magnetodrag, $\rho_{{\rm d},\parallel}$, obtained by summing E-mechanism and P-mechanism contributions with weighting factor $\beta=0.3$, see Eqs.(\ref{eq:drag}),(\ref{eq:emechanism}),(\ref{eq:heatflow}) [here $B=0.4\, {\rm T}$]. P-mechanism dominates far
 from DNP, whereas E-mechanism dominates close to DNP creating a unique configuration of nodal lines. The large negative peak of magnetodrag at DNP is a hallmark of the energy-transport mechanism. Parameters used are same as in Fig. \ref{fig2}. (d) Line trace ($n_1 = -n_2$) along dashed line in panel (c) for various values of $B$. (e) Experimental measurement of $\rho_{\rm d,\parallel}$ from Ref. \onlinecite{geim} displaying the behavior at DNP similar to (d).}
\label{fig1}
\end{figure}

\end{widetext}

This behavior 
has been explained by a new drag mechanism, which arises from interlayer energy transfer due to interlayer electron-electron scattering\cite{Song12,Song13}.
Interlayer energy transfer can couple to charge currents, via lateral heat currents and thermopower, 
generating drag. Hereafter we will refer to this effect as ``E-mechanism."  This mechanism plays a key role near DNP since thermopower coupling peaks close to the Dirac point (Fig. \ref{fig1}b).  As a result, E-mechanism is maximized close to DNP. When both E- and P-mechanisms are summed together (Fig. \ref{fig1}c,d), we find that E-mechanism dominates over P-mechanism at DNP. The resulting dependence resembles the recent measurements of colossal negative drag at DNP reported in Ref. \onlinecite{geim}.

\section{Comparison of the E and P mechanisms}

While on a microscopic level the P and E mechanisms both arise from the same electron-electron interactions, the two contributions to drag are associated with very different physical effects: interlayer momentum transfer vs. interlayer energy transfer and long-range lateral energy transport coupled to charge flow. Accordingly, these effects develop 
on very different lengthscales. For the P-mechanism the characteristic lengthscales are on the order of the Fermi wavelength, which makes this mechanism essentially local. In contrast, the E-mechanism originates from lateral energy transport in the electronic system which is highly nonlocal.  As a result, the two mechanisms have been 
described by very different approaches, microscopic for P-mechanism\cite{tse2007,sensarma2010,narozhny2007,peres2011,katsnelson2011,narozhny2} and hydrodynamical for E-mechanism,\cite{Song12,Song13} which reflects the difference in the characteristic lengthscales. 

Here we adopt a different strategy and develop a unified framework capable of describing these mechanisms on an equal footing. 
To tackle the different lengthscales relevant for the E and P mechanisms, 
a suitable multiscale framework is needed. This framework should also account for the peculiar features  of particle and hole dynamics near the Dirac point: 
Electric currents carried by electrons and holes in the same direction generate momentum flow in opposite directions.\cite{Kashuba08,fritz08} The same is true for energy flow due to particle and hole currents. In the presence of a magnetic field, 
the opposite sign of the Lorentz force on electrons and holes makes charge currents strongly coupled to neutral (energy) currents, resulting in a giant Nernst/Ettingshausen effect.\cite{muller08,zuev,wei,checkelsky}

As we will show, this rich behavior is conveniently captured by a simple two-fluid model. In this model,
carriers in the conduction and valence bands are described as separate subsystems coupled by mutual drag, originating from 
carrier-carrier scattering. As we will see, P-mechanism and E-mechanism drag, for both $B=0$ and $B\neq 0$, can be obtained from the same two-fluid model, allowing us to analyze these contributions to drag even-handedly. 
A model of this type was developed a while ago by Gantmakher and Levinson\cite{Gantmakher78} to describe magnetotransport in charge-compensated conductors, and in particular the anomalies in magnetoresistance and Hall resistance arising at nearly equal electron and hole densities. A similar model was used to describe magnetotransport near CN in graphene in Ref.\onlinecite{Abanin11}. As we will see, this model can successfully account for the strong influence of magnetic field on drag near DNP observed in Ref.\onlinecite{geim}.

\begin{figure}
\includegraphics[scale=0.20]{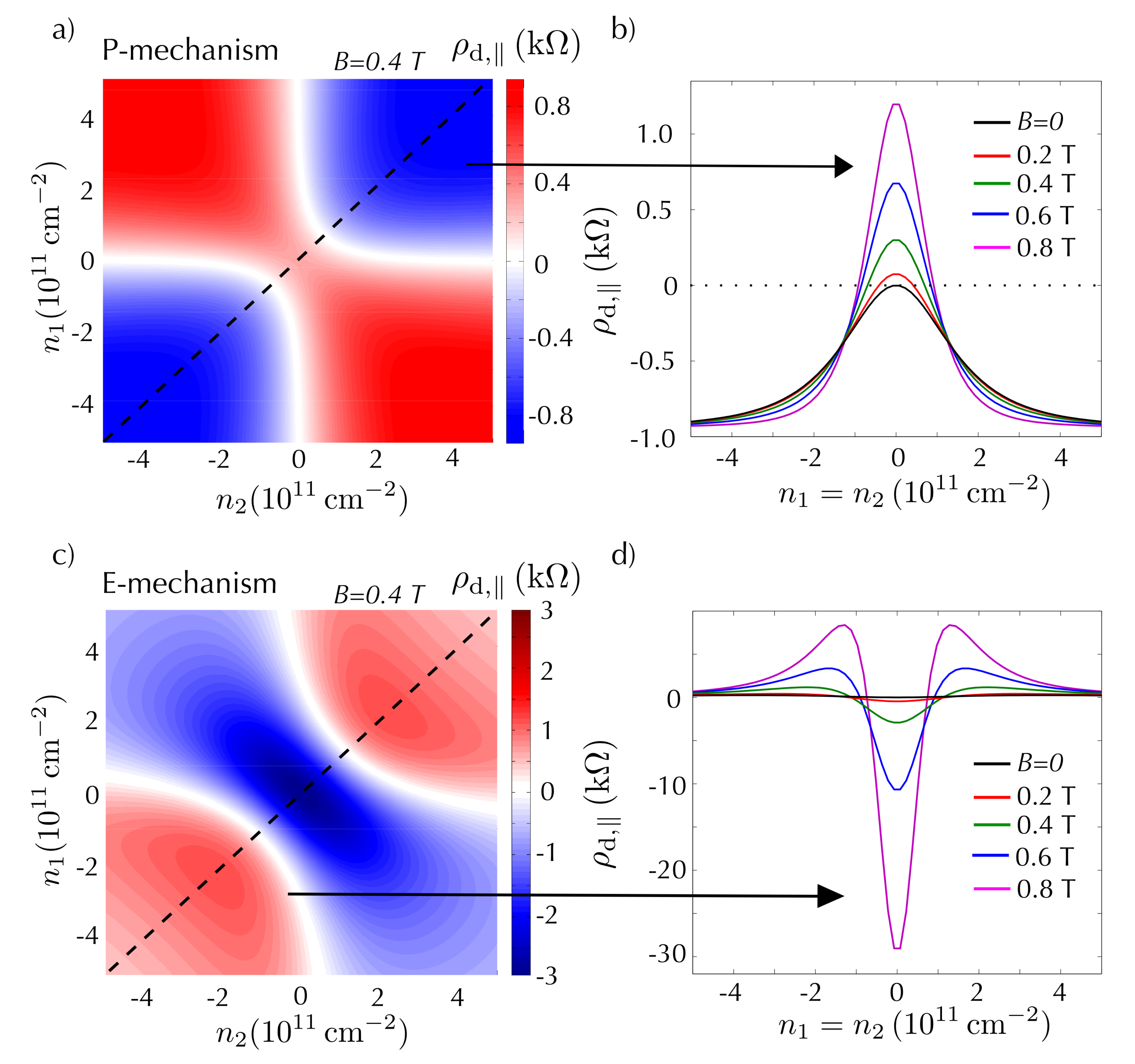}
	\caption{Coulomb drag originating from P and E mechanisms for different values of applied magnetic field. (a)  Magnetodrag density dependence for P-mechanism, obtained from Eq.(\ref{eq:drag}) for $B=0.4\, {\rm T}$, $\eta= 0.23 \hbar$. Note that drag resistance $\rho_{{\rm d}, \parallel}$ is weaker at DNP than away from DNP, and has {\rm positive} sign at DNP. (b) Line traces along dashed line in (a) for various values of $B$. (c) Magnetodrag density dependence for E-mechanism, obtained from Eq.(\ref{eq:emechanism}) at $B=0.4\, {\rm T}$,
 $\eta=0$ and $\beta=1$. Note a large negative peak at DNP which is strongly enhanced by $B$ field, a behavior distinct from that for P-mechanism.
(d) Line traces along dashed line in (c) for various values of $B$.  Disorder broadening of DP of width $\Delta \approx 200 \, {\rm K}$ and $T= 200 \, {\rm K}$ were used here as well as in Figs. \ref{fig1}, \ref{fig3} (see text).
}\label{fig2}
\end{figure}

The two-fluid model predicts
density and magnetic field  dependence which is distinct for the P and E contributions (see Fig. \ref{fig2}). The E contribution features a large peak at DNP, whereas the P contribution is small near DNP and large away from DNP. The peak in the E contribution is sharply enhanced by magnetic field, whereas the P contribution does not show strong field dependence (see Fig. \ref{fig2}). Overall, the density plots for P and E contributions look similar up to an overall sign reversal. This behavior makes it easy to distinguish these contributions in experiment. 

We find that the magnitude of the E contribution can exceed the P contribution as is evident near DNP. Indeed, adding the two contributions up in Fig. \ref{fig1}c,d produces a distinct density dependence of $\rho_{\rm d, \parallel}$ (Fig. \ref{fig1}c). While the exact arrangement of nodal lines can somewhat depend on the parameters chosen, the qualitative features - E-mechanism dominates near DNP (negative $\rho_{\rm d, \parallel}$), and P-mechanism dominates far away from DNP -  are robust.

The general reason for the relative smallness of the P contribution 
can be understood as follows. The two-fluid model  describes coupling between carriers of different types via the mutual drag coefficient $\eta$, see Eq.(\ref{eq:electron}). The dependence of $\eta$ on the interaction strength  $\alpha_0=e^2/\hbar v$
can be obtained\cite{Abanin11} by matching the dependence of conductivity  {\it vs.} $\alpha_0$ found in Refs.\onlinecite{Kashuba08,fritz08}. This gives a general relation of the form
\be
\eta=F(\alpha)\hbar
,\quad
\alpha=\frac{\alpha_0}{\kappa_0+\frac{\pi}{8}N\alpha_0}
,
\ee
where $\alpha$ is the RPA-screened interaction, $N$ is spin/valley degeneracy and $\kappa_0$ is the dielectric constant for the substrate. Since $N=8$ for the double layer system, the factor in the denominator is $\sim 10$ ($\kappa_0\approx 4$ for a BN substrate). A ten-fold reduction of the bare value $\alpha_0\approx 2.4$ yields a small value of RPA-screened interaction, $\alpha\approx 0.25$. The function $F$ admits a power series expansion in $\alpha$, arising from the solution of the quantum Boltzmann equation,\cite{Kashuba08,fritz08} with the leading term being $\alpha^2\sim 0.06$ (which corresponds to the two-particle Born scattering cross-section). This leads to a weak mutual drag, $\eta\ll\hbar$.

Crucially, the E contribution to drag remains unaffected by the small values of $\eta$ so long as the interlayer thermalization occurs faster than the electron-lattice relaxation. This is the case in graphene, since the electron-lattice cooling in this material is dominated by acoustic phonons, giving a slow electron-lattice cooling rate in a wide range of temperatures.\cite{supercollisions,graham2013,betz2013}
As a result, as discussed in more detail in Sec.\ref{sec:Emechanism}, the drag originating from E-mechanism takes on a ``universal value'' which shows little dependence on the interlayer scattering rate. The relative strength of the P and E contributions to magnetodrag at DNP, estimated below, can be characterized by
\be\label{eq:P/E}
\rho_{\rm d,\parallel}^{\rm (P)}\approx -\lp \frac{0.6}{\beta }\frac{\sigma\eta}{e^2} \rp \rho_{\rm d,\parallel}^{\rm (E)}
\ee
where $\sigma$ is the conductivity at charge neutrality and $\beta\sim 1$ is a factor describing temperature gradient buildup in response to energy flow in the system.
Since $\sigma\sim 4e^2/h$ whereas $\eta$ is much smaller than $\hbar$, the factor in parenthesis is much smaller than unity. The smallness of the P contribution, while being quite general, is not entirely universal. In particular, it does not hold  far from DNP, where E-mechanism is small (see Fig.\ref{fig2}). It also does not hold at elevated temperatures when electron-lattice cooling length becomes small compared to system size, $\xi\lesssim W$, leading to small values $\beta\ll 1$ which suppress the E contribution, see Eq.(\ref{eq:beta}). However, near DNP and at not too fast electron-lattice cooling, we expect E-mechanism to overwhelm P-mechanism.

\section{The two-fluid model}
\label{sec:2fluid}

To describe transport near the Dirac point, 
it is crucial to account for the contributions of both electrons and holes. This can be done by
employing the quantum kinetic equation approach of Refs.\onlinecite{Kashuba08,fritz08,Gantmakher78,Abanin11}. For a spatially uniform system, we have
\be\label{eq:kinetic_equation}
q_{e(h)} \lp \vec{E}+\frac{\vec v}{c}\times \vec{B} \rp \frac{\partial f_{e(h)} (\vec{p})}{\partial \vec{p}}=I[f_{e}(\vec{p}), f_h(\vec{p})],
\ee
where $f_{e(h)}(\vec{p})$ is the distribution function for electrons and holes, and $q_e=-q_h=e$. 
The collision integral $I$ describes momentum relaxation due to two-particle collisions and scattering by disorder.
The approach based on Eq.(\ref{eq:kinetic_equation}) is valid in the quasiclassical regime, when particle mean free paths are long compared to wavelength.
This is true for weak disorder and carrier-carrier scattering. Both scattering rates must be smaller than $k_BT/\hbar$,  
which is the case at weak effective fine structure constant $\alpha=e^2/\hbar v_0 \kappa \ll 1$ ($\kappa$ is the dielectric constant).

The kinetic equation (\ref{eq:kinetic_equation}) can be solved analytically in the limit of small $\alpha$.\cite{Kashuba08,fritz08}
Rather than pursuing this route, we will adopt a two-fluid approximation used in Refs.\onlinecite{Gantmakher78,Abanin11} which is particularly well suited for analyzing magnetotransport. In the two-fluid approach, transport coefficients can be obtained from the balance of the net momentum for different groups of carriers, electrons and holes, taken to be moving independently. We use a simple ansatz for particle distribution function, 
\be\label{eq:distributionE}
f_{e(h)}(\vec{p})=\frac{1}{e^{(\epsilon_{\vec{p}}-\vec{p}\vec{a}_{e(h)}- \mu_{e(h)})/k_{\rm B}T}+1}
,\quad
\epsilon_{\vec{p}}=v_0 |\vec{p}|
,
\ee
where $\mu_{e}=-\mu_{h}$ are the chemical potentials of electrons and holes. The quantities $\vec{a}_{e}$ and $\vec{a}_{h}$, which have the dimension of velocity, are introduced to describe a current-carrying state.
This ansatz corresponds to a uniform motion of the electron and hole subsystems, such that the
collision integral for the e-e and h-h processes vanishes (as follows from the explicit form of the collision integral given in Ref.\onlinecite{fritz08}). Thus only the e-h collisions contribute to momentum relaxation, resulting in mutual drag between the e and h subsystems.

In the following analysis we do not account for possible temperature imbalance between electron and hole subsystems since fast e-e and e-h collisions quickly establish thermal equilibrium locally in space. As we will see below, spatial temperature variation across the system becomes essential in the regime dominated by energy transport. We will treat this regime in Sec.\ref{sec:Emechanism}.

Eq.(\ref{eq:kinetic_equation}) yields coupled equations for ensemble-averaged velocities and momenta of different groups of carriers, described by the distribution (\ref{eq:distributionE}):
\footnote{We use the opportunity to correct the sign in the transport equation, Eq.(7) of Ref.\onlinecite{Abanin11}.}
\be\label{eq:electron}
-q_i \lp {\bf E}_i+\vec V_i  \times {\bf B}\rp=-\frac{\vec P_i}{\tau_{i}}-\eta\sum_{i'} n_{i'}   (\vec V_i-{\vec V}_{i'})
,
\ee
where $i,i'=1,2,3,4$ label the e and h subsystems in the two layers. The ensemble-averaged scattering times $\tau_{i}$, the carrier densities $n_i$, and the electron-hole drag coefficient $\eta$, describing collisions between electrons and holes, are specified below. The electric field $\vec E_i$ is the same for electrons and holes in one layer, but is in general different in different layers.

The quantities ${\vec V}_i$,  $\vec P_i$ are
proportional to each other, $\vec{P}_i=m_i \vec{V}_i$. Here the ``effective mass'' is obtained by averaging over the distribution of carriers, as described in Supplementary Information.
The integrals over $\vec p$, evaluated numerically, give the effective mass as a function of $T$ and $\mu$. At charge neutrality, setting $\mu_{e(h)}=0$, we find 
\be
m=\frac{9\zeta(3)}{2\zeta(2)} \frac{k_{\rm B}T}{v_0^2}\approx  3.288 \frac{k_{\rm B}T}{v_0^2}
.
\ee
At high doping, $\mu\gg k_{\rm B}T$, the effective mass is given by the familiar expression, $m=\mu/v_0^2$. In Sec.\ref{sec:Pmechanism}, we will use the approach outlined above to describe momentum drag.

The two-fluid model can also be used to describe energy transport. Indeed, particle flow
is accompanied by heat flow, described by
\be\label{eq:jq}
\vec j_{\rm q}= TS_e n_e\vec V_e+ TS_h n_h \vec V_h
\ee
where $S_e$ and $S_h$ is the entropy per carrier for electrons and holes. Here the entropy and particle density can be related to the distribution function via
\bea\label{eq:S/n}
S_i&=&-\frac{4k_{\rm B} }{n_i}\int\frac{d^2p}{(2\pi)^2}\lb(1- f_i(\vec p)) \ln (1-f_i(\vec p))\right.
\\ \nonumber
&&
\left. + f_i(\vec p)\ln f_i(\vec p)\rb
,\quad n_i=4\int\frac{d^2p}{(2\pi)^2}f_i(\vec p)
.
\eea
In our analysis, we will need the value at charge neutrality. Direct numerical integration in Eq.(\ref{eq:S/n}) gives  $S\approx 3.288 k_{\rm B} $. In Sec.\ref{sec:Emechanism}, we will connect $\vec j_{\rm q}$ to electric current, which will lead to a simple model for drag originating from E-mechanism.

\section{Momentum drag mechanism}
\label{sec:Pmechanism}
Here we will use the two-fluid model introduced in Sec.\ref{sec:2fluid} to derive momentum drag.
To facilitate the analysis of transport equations, it is convenient to switch from vector notation to a more concise complex-variable notation. We will describe velocity, momentum and electric field by complex variables,
\be
\tilde V=V_x+iV_y
,\quad
\tilde P=P_x+iP_y
,\quad
\tilde E=E_x+i E_y
.
\ee

The solution of Eq. (\ref{eq:electron}) can be written in a compact form by introducing the complex-valued quantities
\be
\lambda_i=\frac{n_i}{\frac{m_i}{\tau_i}-iq_iB+\eta N}
,\quad
N=\sum_{i'=1...4}n_{i'}
.
\ee

Solving the transport equations and summing electron and hole contributions to the electric current in each layer 
we 
obtain the current-field relation for the two layers using a $2\times 2$ matrix that couples variables in layer 1 and layer 2:
\be\label{eq:matrix}
\lp\begin{array}{c} \tilde j_1\\ \tilde j_2\end{array}\rp= \lp\begin{array}{cc}\sigma_{11} &\sigma_{12}\\ \sigma_{21} &\sigma_{22} \end{array}\rp
\lp\begin{array}{c} \tilde E_1\\ \tilde E_2\end{array}\rp
.
\ee
Here $\sigma_{11}=e^2\lb (\lambda_{1e}-\lambda_{1h})f_1+\lambda_{1e}+\lambda_{1h} \rb$, $\sigma_{12}=\sigma_{21}= e^2(\lambda_{1e}-\lambda_{1h})f_2$, $\sigma_{22}= e^2\lb  (\lambda_{2e}-\lambda_{2h})f_2+\lambda_{2e}+\lambda_{2h}\rb$, 
\be
f_1=\frac{\eta (\lambda_{1e}-\lambda_{1h})}{1-\eta\Lambda}
,\quad 
f_2=\frac{\eta (\lambda_{2e}-\lambda_{2h})}{1-\eta\Lambda}
.
\ee

Here the quantities $\sigma_{11}$ and $\sigma_{22}$ describe the conductivity of layers 1 and 2, whereas the quantities $\sigma_{12}$ and $\sigma_{21}$ describe mutual drag between the layers (we note that $\sigma_{12}=\sigma_{21}$). The real and imaginary parts of $\sigma_{12}$ describe the longitudinal and Hall drag.

\begin{figure}
\includegraphics[scale=0.20]{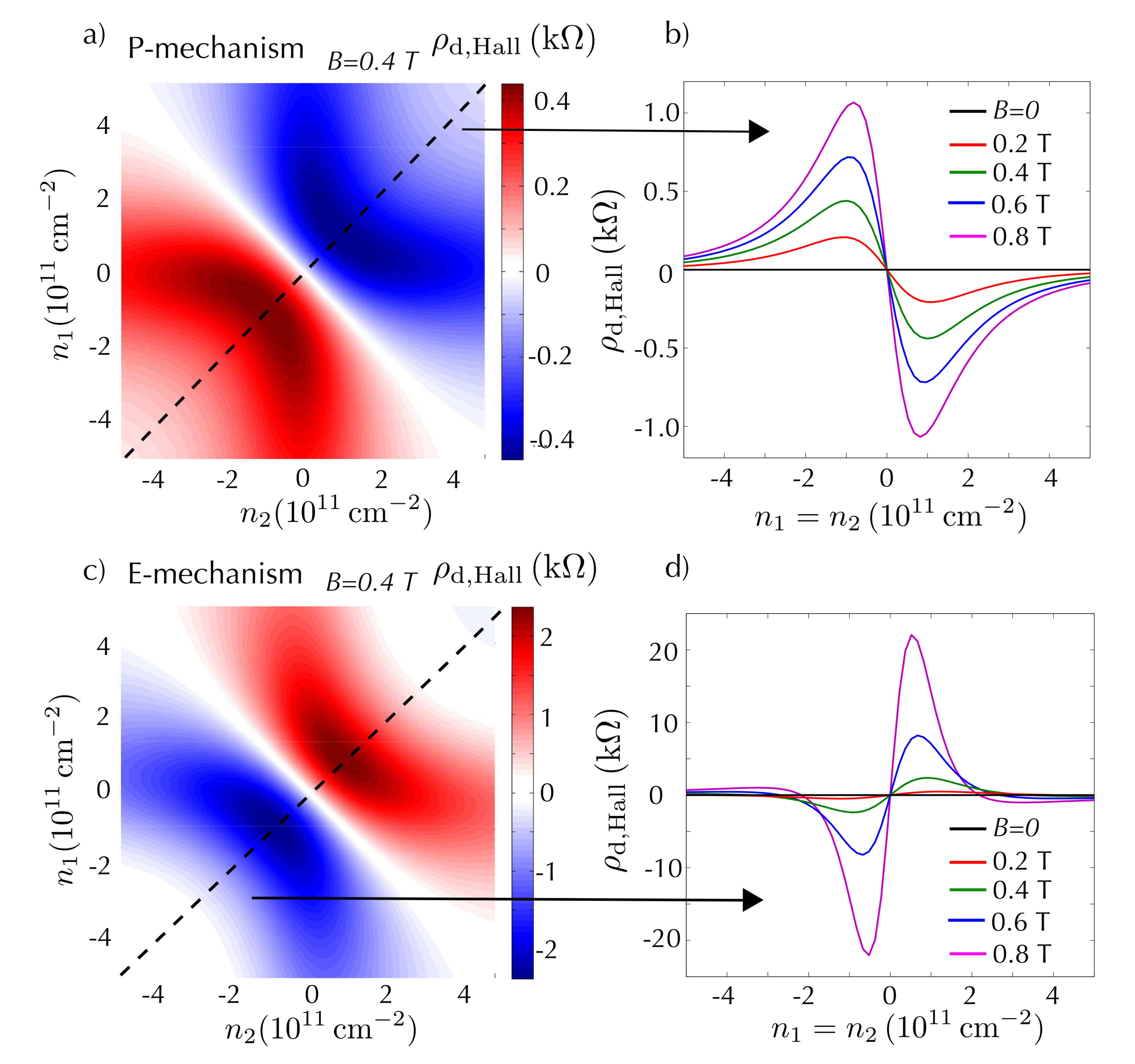}
\caption{Hall drag originating from P and E mechanisms for the same parameters as in Fig.\ref{fig2}. (a)  Hall drag density dependence for P-mechanism, obtained from Eq.(\ref{eq:drag}). (b) Line traces along black dashed line in (a) for various values of $B$. (c) Hall drag density dependence for E-mechanism , obtained from Eq.(\ref{eq:emechanism}). (d) Line traces along black dashed line in (c) for various values of $B$. The difference in sign for the two contributions makes them easy to identify experimentally. Parameters used identical to Fig. \ref{fig2}.
}
\label{fig3}
\end{figure}

The longitudinal and transverse drag resistivity can be obtained by inverting the matrix, Eq.(\ref{eq:matrix}), giving
\be
\rho_{\rm d}=\rho_{\rm d,\parallel}+i\rho_{\rm d, Hall} =-\frac{\sigma_{12}}{\sigma_{11}\sigma_{22}-\sigma_{12}\sigma_{21}}
.
\label{eq:drag}
\ee
The quantities $\rho_{\rm d,\parallel}$ and $\rho_{\rm d, Hall}$ give the magnetodrag and Hall drag shown in Figs.\ref{fig2},\ref{fig3}. 
This quantity features an interesting dependence on carrier density and magnetic field. We will analyze the limit of small $\eta$ (weak interactions). In this case,  we have
$\lambda_i\approx n_i/(\frac{m_i}{\tau_i}-iq_iB)$. This gives the drag resistance
\be\label{eq:drag_weak_eta}
\rho_{\rm d}\approx -\eta\frac{(\lambda_{1e}-\lambda_{1h})(\lambda_{2e}-\lambda_{2h})}{e^2(\lambda_{1e}+\lambda_{1h})(\lambda_{2e}+\lambda_{2h})}
.
\ee
For $B=0$, this quantity vanishes at DNP, $n_{1e}=n_{1h}$, $n_{2e}=n_{2h}$. Drag is negative for equal-polarity doping and positive for opposite-polarity doping, reproducing 
behavior well-known for momentum drag.  

For nonzero $B$,  the Hall drag  and magnetodrag can be obtained 
by expanding
$\Im\lambda_i(B)=\lambda_i(0)(q_i\tau_i/m_i)B +O(B^3)$ in Eq.(\ref{eq:drag_weak_eta}). This gives Hall drag that vanishes exactly at DNP but is nonzero near DNP. For electron and hole densities near DNP, such that $\lambda_e\approx\lambda_h$, we find
\be
\rho_{\rm d, Hall} = -\frac{\eta\tau}{em}B\lp \frac{ (\lambda_{1e}-\lambda_{1h})}{(\lambda_{1e}+\lambda_{1h})}+\frac{ (\lambda_{2e}-\lambda_{2h})}{(\lambda_{2e}+\lambda_{2h})}\rp
+O(B^3)
.
\ee
This expression vanishes on the line $n_1=-n_2$ corresponding to doping of opposite polarity in the two layers.

In contrast to Hall drag,
magnetodrag is nonzero at DNP. From Eq.(\ref{eq:drag_weak_eta}) we obtain a finite magnetodrag of a {\it positive sign:} 
\be
\rho_{\rm d, \parallel} = \eta\frac{\tau^2 B^2}{m^2}+O(B^4)
.
\ee
Here the quantities $\tau$ and $m$ are evaluated at charge neutrality, $n_e=n_h$, in each layer.  Interestingly, the magnetodrag sign comes out opposite to the sign predicted by the energy transport model (see below). The magnetodrag sign therefore provides a clear signature which discriminates between the E and P mechanisms in experiments.

The density dependence of magnetodrag and Hall drag predicted from P-mechanism is shown in Figs.\ref{fig2},\ref{fig3}. In agreement with the above analysis, $\rho_{{\rm d},\parallel}$ in Fig. \ref{fig2}a,b is positive at DNP, increasing quadratically with B field. Also, $\rho_{\rm d, Hall}$ in Fig. \ref{fig3}a,b
increases linearly with $B$ field vanishing along $n_1=-n_2$ as expected.

The plots were obtained by numerically evaluating the expression in Eq.(\ref{eq:drag}), using parameter values described in Fig. \ref{fig2} caption. In all our plots, we found it convenient to account for thermal and disorder broadening of the Dirac point in the same way by setting an effective temperature $T_{\rm eff} = T + \Delta$ in the evaluation of mass and entropy per carrier. We chose a disorder broadening $\Delta = 200 \, {\rm K}$ 
 that corresponds to a Dirac point width $\Delta n \approx 5 \times 10^{10} {\rm cm^{-2}}$ seen in the ultra-clean G/hBN/G devices used for drag measurements\cite{geim}. For simplicity, we also set the scattering rate at neutrality $\tau^{-1}(\mu=0,T=0) = \Delta/\hbar$ [see Supplementary Information for 
further discussion].

We parenthetically note that, while this model reproduces the qualitative features of P-mechanism, it is only valid not too far from DNP. In particular, we have ignored screening which becomes important far away from the Dirac point. As a result, P-drag seen in Figs. \ref{fig2},\ref{fig3} does not diminish with doping. Accounting for screening of the interlayer interaction would generate suppression with doping, in agreement with previous studies of P-drag.\cite{tse2007,sensarma2010,narozhny2007,peres2011,katsnelson2011,narozhny2}

\section{Energy-driven drag mechanism}
\label{sec:Emechanism}

Here we analyze the contribution to drag  resulting from energy transport (E-drag). We will start with evaluating 
the heat current $\vec j_{\rm q}$ [Eq.(\ref{eq:jq})] transported by electric current. In doing so, it will be instructive to first ignore the mutual drag effect discussed above, setting $\eta=0$, and restore finite $\eta$ later. 
Continuing to use complex variables for velocities and fields, we find
$\tilde V_e=(\lambda_eq_e/n_e)\tilde E$, $\tilde V_h=(\lambda_hq_h/n_h)\tilde E$. 
Combining with Eq.(\ref{eq:jq}), we find a relation
\be\label{eq:Q}
\tilde j_{\rm q}=Q\tilde j
,\quad
Q=\frac{T\lp S_e \lambda_e q_e+S_h \lambda_hq_h\rp }{ \lambda_e q_e^2+ \lambda_hq_h^2},
\ee
where $S_{e(h)}$ can be evaluated using the expression in Eq. \ref{eq:S/n}. This relation is particularly useful since the effect of the Lorentz force is fully accounted for via  $\lambda_i$. The imaginary part of $Q$ 
describes the angle between the angle between the heat current and electric current, $\tilde j_{\rm q}$ and $\tilde j$, which corresponds to the Nernst/Ettingshausen effect.

Energy transport, described by Eqs.(\ref{eq:jq}),(\ref{eq:Q}), 
creates temperature gradient across the system. For two layers in close proximity, fast heat transfer between layers due to  interlayer electron scattering leads to a temperature profile which is essentially identical in both layers.\cite{Song12} The temperature gradients can drive a local thermopower via
\be\label{eq:TP}
E=
\frac{Q}{T}\nabla T
,
\ee
where the quantity $Q$ is given by the ratio of the heat current and electric current for the layer in question. As discussed in detail in Ref.\onlinecite{Song13} 
this relation follows from Onsager reciprocity combined with Eq.(\ref{eq:Q}). 

The temperature gradient can be found from balancing the heat flux due to thermal conductivity against the net heat flux due to electric current in the two layers, $j_{\rm q}=j_{\rm 1,q}+j_{\rm 2,q}$. While the details of the analysis somewhat depend on sample geometry (see Ref.\onlinecite{Song13} and discussion below), here we adopt a  simplistic viewpoint and write the balance condition in a general algebraic form as
\be\label{eq:heatflow}
(\kappa_1+\kappa_2)\nabla T=\beta j_{\rm q}
,
\ee
 where $\nabla T$ is the average temperature gradient across the system, $\kappa_1$ and $\kappa_2$ are thermal conductivities of the layers. The quantity $0<\beta\le 1$ is introduced to account for the ``active part'' of the heat that is not lost to contacts and/or the crystal lattice.  

We will first discuss the general behavior that can be understood directly from Eq.(\ref{eq:heatflow}) without specifying $\beta$ value. Combining Eq.(\ref{eq:heatflow}) and Eq.(\ref{eq:TP}) to evaluate drag voltage, we can write drag resistivity as
\be
\rho_{12}=\beta\frac{Q_1 Q_2}{T(\kappa_1+\kappa_2)}.
\label{eq:emechanism}
\ee
This quantity
is symmetric under interchanging layers, $1\leftrightarrow 2$. The real and imaginary parts of $\rho_{12}$ describe magnetodrag and Hall drag. These quantities feature interesting dependence on carrier density shown in Fig.\ref{fig2}c,d and Fig.\ref{fig3} c,d [see 
Fig. \ref{fig2} caption for parameter values]. 
Notably, the signs of magnetodrag and Hall drag obtained from E-mechanism are opposite to those obtained from P-mechanism. 
The relation between the signs of the E and P contributions provides a convenient way to differentiate between the two mechanisms in experiment. 

In our numerical simulations of E-mechanism we used $\eta =0$ to reflect the ``universal values" of drag that E-mechanism takes on in the weak coupling regime. For finite but small $\eta$, E-mechanism remains unaffected. We note, however, that at strong coupling, large values of $\eta$ can affect the magnitude of E-mechanism drag.

The behavior of drag, described by Eq.(\ref{eq:emechanism}),
is particularly simple at charge neutrality. In this case, since $n_e=n_h$, $S_e=S_h$, the particle and hole  contributions to the heat current $j_{\rm q}$ are of equal magnitude. Also, since $\lambda_e=\lambda_h$ at $B=0$, drag resistivity vanishes at zero magnetic field $B$. Furthermore, at finite $B$ the quantities $\lambda_e$ and $\lambda_h$ acquire a relative phase difference, such that $\lambda_e=\lambda_h^*$. As a result, the quantities $Q_1$ and  $Q_2$ that enter Eq.(\ref{eq:emechanism}) are purely imaginary, producing 
drag resistivity that has a negative sign for nonzero $B$. 
 We can obtain magnetodrag by expanding in small $B$, which gives
\be
\rho_{\rm d,\parallel}=-\beta\frac{TS^2}{2\kappa}\lp\frac{\tau}{m}B\rp^2
,
\ee
where $\tau$, $m$ and $\kappa$ are evaluated at charge neutrality. For an estimate, we will relate thermal conductivity to electrical conductivity using the Wiedemann-Franz relation, $\kappa=\frac{\pi^2k_{\rm B}^2T}{3e^2}\sigma$. This relation is valid for degenerate Fermi systems, however we expect it to be also approximately valid near charge neutrality. This gives
\be
\rho_{\rm d,\parallel}\approx -\beta\frac{3S^2e^2}{2\pi^2\sigma k_{\rm B}^2}\lp\frac{\tau}{m}B\rp^2
.
\ee
Comparing to the answer for P-mechanism, we find the ratio of the  contributions due to momentum and energy mechanisms
\be
\frac{\rho_{\rm d,\parallel}^{\rm (P)}}{\rho_{\rm d,\parallel}^{\rm (E)}}=-\frac{2\pi^2 k_{\rm B}^2}{3\beta S^2}\frac{\sigma\eta}{e^2}
.
\ee
We can estimate entropy per carrier at DP by evaluating the integral over energy in Eq.(\ref{eq:S/n}). Using the value $S\approx 3.288 k_{\rm B} $ quoted above, we arrive at Eq.(\ref{eq:P/E}). 
Given the conductivity value at charge neutrality, $\sigma\approx 4e^2/h$, and taking into account that the mutual drag coefficient $\eta$ is small when the fine structure constant $\alpha=e^2/\hbar v\kappa$ is small,\cite{Kashuba08,fritz08} $\eta\sim\alpha^2$, we conclude that the ratio in Eq.(\ref{eq:P/E}) is smaller than unity. This indicates that under very general conditions the E contribution overwhelms the P contribution in the DNP region.

The value of $\beta$ in Eq.(\ref{eq:heatflow}) depends on the rate of heat loss from electrons to the lattice and contacts. As an illustration, we consider the case when heat loss is dominated by cooling to the lattice. 
In this case, $\beta$ depends on the relation between electron-lattice cooling length and system dimensions. 
We can model heat transport across the system as 
\be
\kappa (- \nabla^2  + \xi^{-2}) T = - \nabla j_{\rm q},\quad
0<x<W,
\ee 
where $W$ is system width and $\xi$ is the electron-lattice cooling length. Spatially uniform heat current $j_{\rm q}$ translates into a pair of delta-function sources, localized at $x=0$ and $x=W$. Solving for the temperature profile, we obtain 
the temperature imbalance sustained between the sample edges, $\Delta T = T_{x=W} - T_{x=0} = \beta W j_{\rm q}/\kappa$, with the $\beta$ value given by
\be\label{eq:beta}
\beta = \frac{1}{c} {\rm tanh} c , \quad c = \frac{W}{2 \xi}.
\ee 
This gives $\beta\to1$ when $\xi\gg W$ (slow cooling) and $\beta\to0$ when $\xi\ll W$ (fast cooling). The cooling length, $\xi$, in graphene can be as large as several microns for a wide range of temperatures up to room temperature.\cite{supercollisions,graham2013,betz2013} For such temperatures, since in typical devices $W$ is a few microns or smaller, the factor $\beta$ can be close to unity, $\beta \sim 1$. However, the electron-lattice cooling rate grows at temperatures exceeding a few hundred kelvin owing to cooling pathway mediated by optical phonons. At such high temperatures, since the cooling length $\xi$ shortens rapidly, Eq.(\ref{eq:beta}) predicts vanishingly small $\beta$. In this case,  temperature gradients in the electron system do not build up, rendering the E-mechanism ineffective. The latter regime (fast cooling) is not relevant, however, for practically interesting temperatures $T \lesssim 300 \, {\rm K}$, where we expect $\beta\sim 1$ for few-micron-size devices.

We also note that $\beta$ may be altered in a nontrivial way by boundary conditions, for example by contacts that act as heat sinks. In particular, in anisotropic systems or in systems with anisotropic contact placement, the relation between heat flow and $\nabla T$ can become anisotropic. In this case,  $\beta$ can be described as a $2\times2$  tensor (see Supplementary Information for 
further discussion). While the qualitative behavior discussed above (drag order of magnitude and sign at DNP) is not expected  to be altered by anisotropy in heat loss, the tensor character of $\beta$ can affect the layer symmetry of the resultant drag [e.g. see Ref.\onlinecite{Song13} where the lack of symmetry $n_1\leftrightarrow n_2$ stems from anisotropic device geometry].  
In contrast, in the isotropic case, where heat flow is not influenced by device geometry or contact placement, $\beta$ is a c-number. In this case, Eq.(\ref{eq:emechanism}) predicts drag obeying layer symmetry,  $n_1\leftrightarrow n_2$.

Finally, we comment on the anomalously large values of $\rho_{\rm d, \parallel}$ at the highest $B$ fields seen in Fig. \ref{fig2}(d). These values far exceed P-mechanism, however they also exceed the in-plane sheet resistivity. This signals that our treatment, while successfully capturing E-mechanism drag for low $B$, ceases to be valid for higher $B$.  We note in this regard that for $B=0.8 \, {\rm T}$ the energy of the first Landau level, $E_1 \approx 380 \, {\rm K} $, 
exceeds our disorder broadening value $\Delta = 200 \, {\rm K}$.
This hints at the importance of Quantum Hall physics at such fields.

\section{Summary}
In summary, we argue that drag in graphene near charge neutrality is dominated by energy transport effects (E-mechanism) arising due to fast interlayer energy relaxation that couples to lateral energy flow and, via thermopower, drives electric current.  We developed a two-fluid framework which accounts both for the E-mechanism as well as for the standard momentum-transfer drag (P-mechanism), capturing the essential features of the two mechanisms. This unified approach is particularly instructive, not only because it produces both P-mechanism and E-mechanism, but also because it allows an unbiased way of comparing the magnitudes of the two mechanisms. Strikingly, the P and E mechanisms yield opposite sign for both magnetodrag and Hall drag resistivities. Along with a strong peak in magnetodrag at DNP originating from E-mechanism, this sign difference provides a clear way to experimentally distinguish the two mechanisms.

We show that the magnitude of drag originating from the two mechanisms is dominated by very different effects. The P mechanism is mostly controlled by the interlayer electron-electron interaction, becoming weak when this interaction decreases due to large layer separation or screening. In contrast, the E mechanism is controlled by long-range energy transport, yielding a ``universal value'' for drag: the E contribution is essentially independent of the interlayer carrier scattering rate so long as it is faster than electron-lattice cooling. low electron-lattice cooling in graphene ensures that drag near DNP can remain large even when interlayer electron interactions are weak.
This makes graphene an ideal system to observe  E-drag and thereby probe energy transport on the nanoscale.   

We acknowledge useful discussions with A. K. Geim, P. Jarillo-Herrero, L. A. Ponomarenko, and financial sup-
port from the NSS program, Singapore (JS).

Upon completion of this manuscript, we became aware of a related work by Titov et al.\cite{titov2013}

\section{Supplementary Information: Modelling procedure}
\label{sec:modelling}

Here we comment on the quantities that enter the two-fluid description, and discuss the sensitivity of the results to the simplifying assumptions made in the model. 

In the two-fluid model we describe the momentum-velocity relation for each component as $\vec P=m_i\vec V$, where $m_i$ is an ``effective mass.''
An explicit expression for $m_i$ as a function of $T$, $\mu$ can be found by expanding the distribution functions [Eq.(4) of the main text]
to lowest non-vanishing order in $\vec{a}_{e(h)}$:
\be\label{eq:eff_mass}
m_i = \frac{1}{v_0}
\frac{\int d^2\vec p\, p_x \nabla_{\vec a_x}f_i(\vec p)}{\int d^2\vec p\, \frac{p_x}{p}\nabla_{\vec a_x}f_i(\vec p)}
=\frac{1}{v_0}
\frac{\int d^2\vec p\, p_x^2 g_i(\vec p)}{\int d^2\vec p\, \frac{p_x^2}{p}g_i(\vec p)}
,\quad
\ee
where $g_i(\vec p)=f_i(\vec p)(1-f_i(\vec p))$.

The times $\tau_{i}$ for disorder scattering and carrier densities $n_i$ in
Eq.(5) of the main text are expressed through the distribution function [Eq. (4) of the main text]
with $\vec a_i=0$:
\be\label{eq:tau_ensem_ave}
\frac1{\tau_{i}}=\frac4{n_i} \int\!\!\frac{d^2\vec{p}}{(2\pi)^2} \frac{ f_i(\vec{p})}{\tau_{i}(\epsilon_{\vec{p}})}
,\quad
n_i= 4\int\!\!\frac{d^2\vec{p}}{(2\pi)^2}  f_i(\vec{p})
,
\ee
where the factor of four accounts for spin-valley degeneracy in each layer. We pick a model for the transport scattering time $\tau(\epsilon)$ to account for the experimentally observed linear dependence of conductivity vs. doping, $\sigma=\mu_* |n|$, where $\mu_*$ is the mobility away from the DP. This is the case for Coulomb impurities or strong point-like defects, such as adatoms or vacancies\cite{CastroNeto09}. 
In both cases the scattering time has an approximately linear dependence on particle energy,
\be\label{eq:scat_time}
\tau(\epsilon)_{|\epsilon|\gtrsim \gamma}=\hbar |\epsilon|/\gamma^2
, \quad
\gamma=v_0\sqrt{e\hbar/\mu_*}
\ee
where the disorder strength parameter $\gamma$ is expressed through mobility. The value $\mu_*=6\cdot 10^4\,{\rm cm^2/V\cdot s}$ measured in graphene on BN~\cite{Dean10} yields $\gamma\approx 120\,{\rm K}$. Similar values for $\gamma$ are obtained from the  DP width extracted from the resistivity density dependence,\cite{geim} $\Delta n \approx  10^{10}\, {\rm cm^{-2}}$.

In doing simulations, we found it convenient to use
a different, 
simplified model for transport scattering which does not involve integration over particle distribution, 
yet yields results similar to those obtained from a more microscopic model, Eq.(\ref{eq:tau_ensem_ave}). We model the scattering time in the full range of doping densities as
\be
\tau_i = (m_i v_0^2+\Delta)\frac{\hbar }{\Delta^2}
\label{eq:tau}
\ee
where $m_i$ depends on temperature and density via Eq.(\ref{eq:eff_mass}), and the parameter $\Delta$ describes the smearing of DP due to disorder.
This model accounts for the experimentally observed linear dependence of conductivity vs. doping. In the simulation we used the value $\Delta=200\,{\rm K}$ which translates into DP width of about $\Delta n \approx 5\times 10^{10}\, {\rm cm^{-2}}$, consistent with the above estimates. 

Additionally, we found it convenient to account for disorder broadening of DP by using an effective temperature $T_{\rm eff} = T + \Delta$ in the evaluation of the effective mass in Eq.(\ref{eq:eff_mass}) and the entropy per particle in Eq.(8) of the main text. This simple procedure captures the essential characteristics of DP broadening since smearing of the density of states by temperature and disorder occur in a similar fashion.

Using the parameters $\beta=1$, $\Delta = 200 \, {\rm K}$, $T=200 \, {\rm K}$ and the model for scattering time $\tau_i$ in Eq.(\ref{eq:tau}), we plot $\rho_{{\rm d}, \parallel}$ and $\rho_{\rm d, Hall}$ for P-mechanism [Eq.(13) of the main text and using $\eta=0.23 \hbar$] and E-mechanism [Eq.(20) of the main text and using $\eta = 0$] in Figs.2,3 respectively. As discussed above, this gives  density dependence of drag that differs in sign for the two mechanisms. Additionally, we find that the E-mechanism magnitude exceeds that of P-mechanism for the region near DNP. This agrees with the small ratio for P-mechanism vs. E-mechanism derived for small $\eta$.

\section{Supplementary Information: Tensor $\beta$}
Since E-mechanism depends on long-range energy transport, it is particularly sensitive to sample geometry and the arrangement of contacts. As discussed in Ref. \onlinecite{Song13}, the source and drain contacts  
can act as ideal heat sinks
that suppress temperature gradients along current flow. In contrast, for a Hall bar with ``non-invasive''  voltage probes, temperature gradient in the direction transverse to current flow will remain essentially unaffected by the voltage probes. This anisotropy can be easily implemented by making $\beta$ in Eq. (19) of the main text a tensor, 
\be\label{eq:beta_general}
\beta=\lp\begin{array}{cc} \beta_{xx} & \beta_{xy}\\ \beta_{yx} & \beta_{yy}\end{array}\rp
\ee
so that heat current in one direction generates higher temperature gradient than heat current in another direction. For example, the Hall bar geometry analyzed in Ref. \onlinecite{Song13} is described by a tensor with eigenvalues approximately equal 0 and 1 for the directions along and transverse to the Hall bar, respectively. Choosing the  $x$ axis along the bar, we obtain
\be\label{eq:beta01}
\beta=\lp\begin{array}{cc} 0 & 0\\ 0 & 1\end{array}\rp
\ee
Plugging this matrix in Eq. (19) of the main text
and proceeding as in the main text, we arrive at an expression for drag resistivity
\be
\rho_{12}=\frac{i Q_2{\rm Im}(Q_1)}{T(\kappa_1+\kappa_2)}, 
\label{eq:Qbeta01}
\ee 
The dependence on density can be obtained by plugging 
$Q$ from Eq. (17) of the main text
into Eq.(\ref{eq:Qbeta01}) of the Supplement. This gives $\rho_{\rm d,\parallel}$ and  $\rho_{\rm d,Hall}$ maps shown in Fig. \ref{fig3}a,b. Note the absence of layer symmetry  $1 \leftrightarrow 2$ which was present in the expression (20) of the main text 
and manifest in drag maps in Figs.1,2 in the main text. The density dependence shown in Fig.\ref{fig3} is essentially identical to that found in Ref.\onlinecite{Song13}.

\begin{figure}
\includegraphics[scale=0.20]{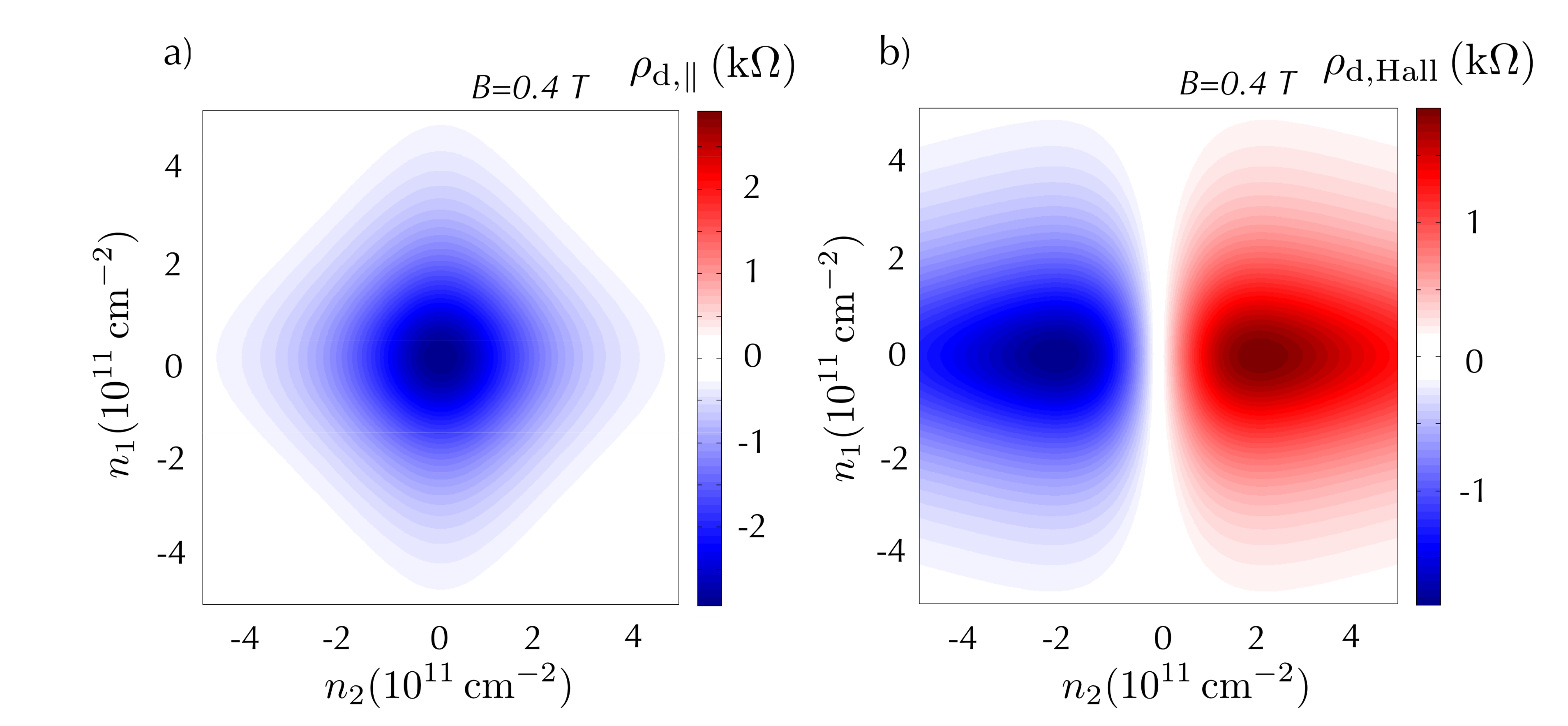}
\caption{Long-range energy flow which governs E-mechanism leads to anisotropy in the quantity $\beta$ which relates heat flow and temperature gradient, Eq.(\ref{eq:beta_general}). The effect of system anisotropy on drag originating from E-mechanism is illustrated for $\beta$ given in Eq.(\ref{eq:beta01}). Density dependence of $\rho_{{\rm d}, \parallel}$ and $\rho_{{\rm d}, {\rm Hall}}$, obtained from Eq.(\ref{eq:Qbeta01}), is shown for $B=0.4\, {\rm T}$ using parameter values identical to those in Figs. 2,3 of the main text.
}
\label{fig3}
\end{figure}

\end{document}